\documentclass[12pt,preprint]{aastex}

\usepackage{natbib}  
\usepackage{emulateapj5}

\slugcomment{}

\shorttitle{Butterfly Star: Submillimeter Structure}
\shortauthors{Wolf et al.}

\newcommand{\about}    {$\sim$\ts}
\newcommand{\aboutless}{$\simless$\ts}

\newcommand{\ts}{\thinspace}
\newcommand{\simless}{\mathbin{\lower 3pt\hbox
     {$\rlap{\raise 5pt\hbox{$\char'074$}}\mathchar"7218$}}}
\newcommand{\simgreat}{\mathbin{\lower 3pt\hbox
     {$\rlap{\raise 5pt\hbox{$\char'076$}}\mathchar"7218$}}}

\begin{document}


\title{Submillimeter Structure of the Disk of the Butterfly Star}


\author{S.\ Wolf, A.\ Schegerer, H.\ Beuther}
\affil{Max Planck Institute for Astronomy, 
K\"onigstuhl 17, 69117 Heidelberg, Germany}
\email{swolf@mpia.de, schegerer@mpia.de, beuther@mpia.de}

\and

\author{D.\ L.\ Padgett}
\affil{California Institute of Technology, 1200 E California Blvd,
Mail code 220-6, Pasadena, CA 91125}
\email{dlp@ipac.caltech.edu}

\and

\author{K.\ R.\ Stapelfeldt}
\affil{Jet Propulsion Laboratory, California Institute of Technology, 
  4800 Oak Grove Drive, Pasadena, CA 91109}
\email{krs@exoplanet.jpl.nasa.gov}




\begin{abstract}
We present a spatially resolved 894\,$\mu$m map of the circumstellar disk
of the Butterfly star in Taurus (IRAS\,04302+2247),
obtained with the Submillimeter Array (SMA).
The predicted and observed radial brightness profile agree well in the outer disk region,
but differ in the inner region with an outer radius of \about80-120\,AU.
In particular, we find a local minimum of the radial brightness distribution at the center,
which can be explained by an increasing density / optical depth 
combined with the decreasing vertical extent of the disk towards the center.
Our finding indicates that young circumstellar disks can be optically thick
at wavelengths as long as 894\,$\mu$m.
While earlier modeling lead to general conclusions about the global disk structure
and, most importantly, evidence for grain growth in the disk (Wolf, Padgett, \& Stapelfeldt~2003), 
the presented SMA observations provide more detailed constraints for the disk
structure and dust grain properties in the inner, potentially planet-forming region 
(\aboutless80-120\,AU) vs.\ the outer disk region (\about120-300\,AU). 
\end{abstract}




\section{Introduction}\label{intro}

IRAS~04302+2247 is a Class~I protostar in the Taurus-Auriga molecular cloud complex
whose equatorial plane is inclined edge-on to the line of sight
(inclination = 90$^{\rm{o}} \pm 3^{\rm{o}}$; Wolf et al.~2003, hereafter WPS03). 
Parallel to the increasing amount of observational constraints, 
such as ground-based near-infrared images and polarization maps (Lucas \& Roche~1997, 1998),
near-infrared images obtained with the Hubble Space Telescope (Padgett et al.~1999)
and
spatially resolved 1.3\,mm and 2.7\,mm maps (WPS03),
several attempts have been undertaken to model the structure 
and physical conditions in the circumstellar disk and envelope of this object
(e.g., Lucas \& Roche~1998, WPS03, Stark et al.~2006).
In particular, spatially resolved images of the circumstellar environment 
of the Butterfly Star, obtained in the near-infrared and millimeter wavelength range, 
allowed WPS03 to conclude that the grains in the envelope of
this object cannot be distinguished from those of the interstellar medium, 
while grains have grown via coagulation by up to 2 - 3 orders of magnitude 
in the much denser circumstellar disk. 
The separated dust grain evolution is in agreement with the theoretical prediction
of a sensitive dependence of grain growth on the location in the circumstellar environment
of young (proto)stars: Grain growth is expected to occur on much
shorter timescales in the dense region of circumstellar disks than
in the thin circumstellar envelope. 
For the same reason a radial dependence of the dust grain evolution
in the disk itself is expected.
However, the observational data presented by WPS03 did not allow to constrain
the spatial dependence of the dust grain properties in the disk.

Based on radiative transfer simulations, using the disk model by WPS03, we found
that further insights into and constraints for the dust grain growth as the first stage 
of planet formation in the circumstellar disk of the Butterfly Star can be obtained
with high-resolution submillimeter observations.
As outlined in Sect.~\ref{sect.ana}, the apparent structure of the disk is predicted to change significantly
as the observing wavelength is decreased from millimeter to submillimeter wavelengths,
allowing to constrain the radial and vertical disk structure and distribution 
of the dust grain properties.

In this paper we present and discuss new, spatially resolved observations of the circumstellar disk 
of the Butterfly Star, obtained with the Submillimeter Array\footnote{
The Submillimeter Array is a joint project between the Smithsonian Astrophysical Observatory 
and the Academia Sinica Institute of Astronomy and Astrophysics 
and is funded by the Smithsonian Institution and the Academia Sinica.} 
(SMA) at 894\,$\mu$m.
The observations and data reduction are described in Sect.~\ref{sect.obs}, 
followed by a description of the data analysis (Sect.~\ref{sect.ana}) 
and conclusions in Sect.~\ref{sect.sum}.

\section{Observations and Data Reduction}\label{sect.obs}

IRAS\,04302+2247 was observed with the SMA \cite{ho} on January~$9^\mathrm{th}$, 2006, 
using the upper and lower side bands at 330\,GHz and 340\,GHz, respectively.
The phase center was 
R.A.\ $=4^{\rm h}33^{\rm m}16.219^{\rm s}$ and Dec.\ $=22^{\circ}53'20.''29$ (J2000.0). Both,
the upper and lower sideband had a width of 2\,GHz consisting of $24 \times
256$ channels with a spectral resolution of 0.41\,MHz. 
The angular diameter, i.e., the radial extent of the disk
derived from the previous 1.36\,mm continuum observations 
amounts to $\sim$\,$4.''3$ (WPS03).
For this reason we used the extended configuration of
the array with 7 antennas and corresponding projected baseline lengths of 
$24$\,m -- $254$\,m during the observation. The resulting size of the
synthesized beam is $0.''67 \times
0.''53$ with a position angle of $104^{\circ}$. The weather conditions during 
the 6\,hours of observation of the target were stable with typical 225\,GHz zenith sky 
opacities of $\tau=0.05-0.07$ measured by the National Radio Astronomy 
Observatory (NRAO) tipping radiometer operated by the Caltech Submillimeter 
Observatory (CSO). This result corresponds to an opacity of $\tau \approx
0.23$ at 345\,GHz \cite{masson}. The radio galaxy
3c111 was used for gain calibration while passband calibration was done with
the quasar 3c273. The flux calibration was determined with Uranus. 
The SMA data were reduced and calibrated with the MIR software package
\cite{qi}. The 1\,$\sigma$ continuum r.m.s amounts to
3.3\,mJy\,beam$^{-1}$. 

\begin{figure}[h!]
  \begin{center}
    \resizebox{.99\hsize}{!}{\includegraphics[angle=0]{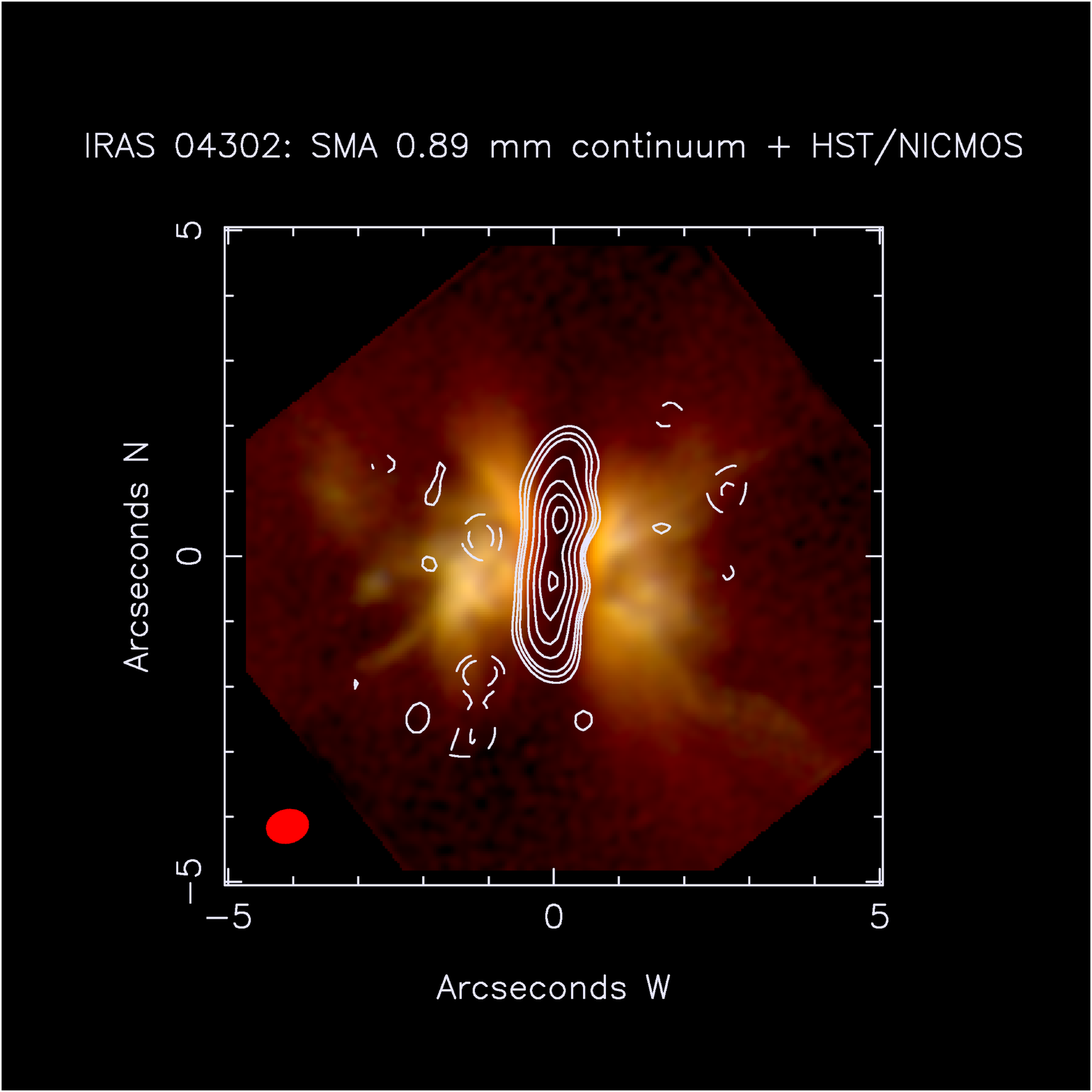}}
  \end{center}
  \caption{
    Submillimeter map of the Butterfly Star (894\,$\mu$m, contour lines),
    overlayed on the near-infrared scattered light map (Padgett et al.~1999).
    The levels of the solid contour are 
    6.8,
    10.2,
    13.7,
    23.9,
    34.1,
    44.3, and
    54.5 mJy/beam
    which correspond to the 
    2$\sigma$, 
    3$\sigma$, 
    4\,$\sigma$, 
    7\,$\sigma$, 
    10\,$\sigma$, 
    13\,$\sigma$, and 
    16\,$\sigma$ levels.
    The levels of the dashed contours are
    -6.8 and
    -10.2 mJy/beam
    which correspond to the -2\,$\sigma$ and -3\,$\sigma$ level, respectively.
  }
  \label{sma+hst.map}
\end{figure}

The resulting 894\,$\mu$m continuum map of the disk of the Butterfly Star is shown 
in Fig.~\ref{sma+hst.map}.
In order to remove the potential influence of the $^{13}$CO(3-2) line at
330.587\,GHz on the total flux and continuum map, we cut a band with a width
of 16\,MHz around its corresponding position in the spectrum.

The 894\,$\mu$m continuum flux of IRAS\,04302+2247
amounts to 267\,mJy with a corresponding calibration error of \about$15\%$.
This flux is consistent with our predicted flux of 225\,mJy.
Consequently, our SMA measurements confirm the previously estimated disk mass
($M_{\rm disk}$=0.07 M$_\sun$; WPS03). 
We wish to emphasize that this flux / mass estimate does not rely 
on the assumption of optically thin emission, but was derived
from the self-consistent model by WPS03 
(i.e., self-consistent with respect to the disk density distribution, 
dust grain parameters, and resulting disk temperature structure; 
see Sect.~\ref{s.sect.model} for details).

\section{Data Analysis}\label{sect.ana}

\subsection{Disk model}\label{s.sect.model}

In the following we provide a brief summary of the disk model
we use in the subsequent data analysis and discussions (see WPS03 for more details):

\noindent
{\em Density profile}:
We assume a disk density profile as described by Shakura \& Sunyaev~(1973):
\begin{equation}\label{dendisk}
\rho_{\rm disk} = \rho_0   
\left( \frac{R_{*}}{\varpi} \right)^{\alpha}
\exp \left\{ -\frac{1}{2} \left[ \frac{z}{h(\varpi)} \right]^2 \right\},
\end{equation}
where $\varpi$ is the radial distance from the star in the disk midplane, $R_{*}$ is the stellar radius,
and $h(\varpi)$ is the disk scale height $h = h_0 (\varpi/R_{*})^{\beta}$.
For the exponents $\alpha$ and $\beta$ which describe the radial density profile and disk flaring
we use the relation $\alpha = 3 (\beta - 1/2)$,
which results from viscous accretion theory (Shakura \& Sunyaev~1973).

In addition to the circumstellar disk, the model for the environment
of IRAS~04302+2247 consists of a rotating, infalling envelope. 
However, based on the analysis of the relative contributions of the disk and envelope
to the spectral energy distribution, the second component has no significant influence
on the 894\,$\mu$m data discussed here.

\noindent
{\em Heating sources}:
The main heating source of the circumstellar environment is the embedded star.
We assume typical parameters of a T~Tauri star:
$R_{*} = 2\,{\rm R}_{\sun}$, $T_{*} = 4000\,{\rm K}$ (Gullbring~1998) which correspond to a luminosity of
$L_{*} = 0.92\,L_{\sun}$ under the assumption of a blackbody.
Further heating of the disk is provided by accretion, where we apply 
the viscous disk model by Lynden-Bell \& Pringle (1974).

\noindent
{\em Dust grain properties:}
We consider the dust grains to be homogeneous spheres (radius $a$) with a size distribution described 
by a power-law of the form $ n(a) \propto a^{-3.5}$.
The dust grain ensemble consists of silicate and graphite grains with relative abundances of
62.5~\% astronomical silicate and 37.5~\% graphite.
We use the optical data
of ``smoothed astronomical silicate'' and graphite published by Weingartner \& Draine~(2001) and Draine \& Lee~(1984).
We assume a gas-to-dust mass ratio of 100:1 and a grain mass density of $2.5{\rm g}\,{\rm cm}^{-3}$.

\noindent
{\em Results:}
Radiative transfer simulations are used to calculate the disk temperature structure
self-consistently and to derive observables subsequently.
These simulations are performed with the three-dimensional 
continuum radiative transfer code MC3D (e.g., Wolf et al.~1999, Wolf~2003).
Based on this model, WPS03 find the following best-fit model parameters:
The outer radius amounts to 300\,AU,
$\beta = 58/45$ ($\alpha=213/90$), 
$h$(100\,AU)=15\,AU, and
$M_{\rm disk} = 0.07\,{\rm M}_{\sun}$.
In particular, an upper grain size of \about$100\,\mu{\rm m}$ was found.

\subsection{Comparison of the model predictions with the submillimeter observation}\label{s.sect.comp}

\begin{figure}[t!]
  \begin{center}
    \resizebox{1.0\hsize}{!}{\includegraphics[angle=00]{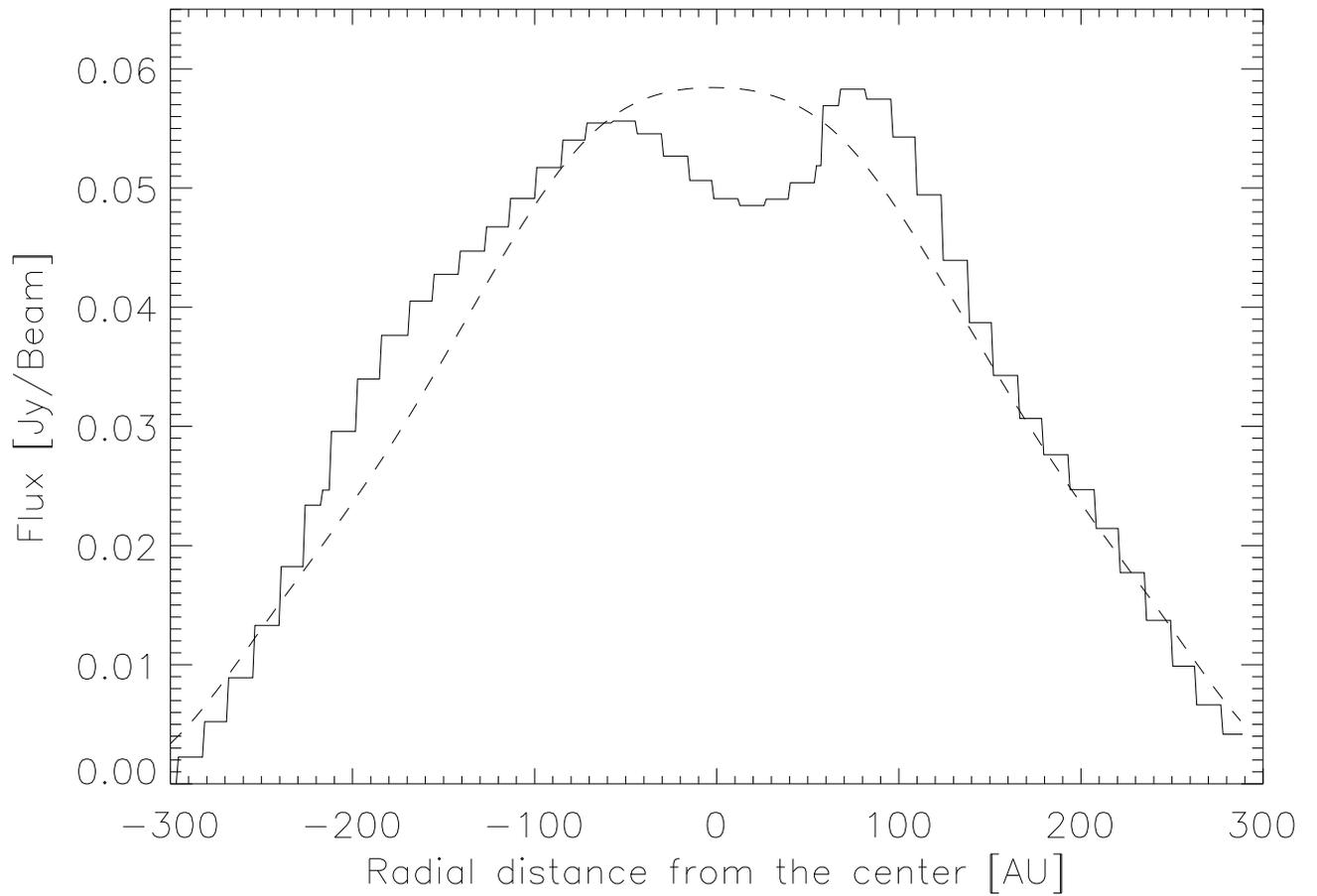}}
  \end{center}
  \caption{    
    Radial brightness profile.
    {\em Solid line:}  Observed 894\,$\mu$m profile.
    The maxima to the left and the right of the center are \about1.4$\sigma$ and \about1.8$\sigma$ 
    above the local minimum, respectively ($\sigma$: standard deviation). 
    The radial scale is based on an assumed distance of 140\,pc.
    {\em Dashed line:} 
    For comparison: 
    Simulated 894\,$\mu$m profile, convolved with the corresponding synthesized SMA beam
    (based on the model by WPS03).
    For comparison with the 1.3\,mm and 2.7\,mm radial brightness profiles see Fig.~3 in WPS03.
  }
  \label{sma.radprof}
\end{figure}

In Fig.~\ref{sma.radprof} the radial brightness profile along the disk midplane is shown (solid line).
It agrees well with the predicted profile (dashed line) in the outer regions of the disk
with radial distances larger than \about80-120\,AU from the center.
Thus, our observations confirm the radial density distribution and dust grain properties
for the outer, cold region of the disk.
Consequently, our observations confirm the disk size (radius: 300 AU) derived by WPS03.

Inside \about80-120\,AU we find a discrepancy between the model prediction
and the observed brightness profile. While the model predicts a rather flat plateau for this region,
the observed map even shows a local minimum at the potential location of the star.
The discrepancy amounts to only \about2$\sigma$. 
However, we want to stress that the location of this mininum is not arbitrary, but at the disk center.
Furthermore, the corresponding local maxima show 
a slight symmetry with respect to the disk center.

Two explanation for this brightness minimum are at hand:
First, the minimum could be caused by the lack of emitting dust, i.e., 
an inner region void of small dust grains.
Inner holes in circumstellar disks in different stages of their evolution
have been inferred from their mid-infrared spectral energy distribution
and spatially resolved images. In the particular case of young disks (as the counterpart
to evolved, debris-type disks), inner disk radii which are much larger than the dust
sublimation radius have been deduced from the mid-infrared spectral energy distribution
for example in the case of TW~Hydrae (Calvet et al.~2002) and GM~Aurigae (Rice et al.~2003).
As radiative transfer simulations based on the disk model by WPS03 show, 
the radius of the dust free region would have to amount to \about70\,AU
(compared to a radius of \about300\,AU of the disk).
We can test the hypothesis of the existence of a large inner hole by comparing 
the corresponding 1.36\,mm simulated images with our continuum observations
we obtained at this wavelength with the Owens Valley Radio Observatory (OVRO; WPS03).
We find that this gap would have caused a similar local minimum at 1.36\,mm
(note that the size of the synthesized beam was similar to that of our SMA observations:
$0.64'' \times 0.52''$).
Since this was not observed (see Fig.~3 in WPS03), we can exclude this explanation.

One usually would apply a second independent test to the above conclusion.
If considering the disk alone, the amount of the optical to mid-infrared flux
depends sensitively on the existence or absence of such a large inner hole.
However, this test is not applicable in our case because the emission
of the Butterfly star is dominated by the circumstellar envelope
up to wavelengths of \about174\,$\mu$m (WPS03).

The second and thus remaining explanation is the lack of ``visible'' emitting dust grains.
In particular, the model predicts an optical depth in the midplane (as seen from the star)
which amounts to $\tau_{894\mu{\rm m}} \approx 2 \times 10^3$.
Going from millimeter to submillimeter wavelengths, the dust in the innermost,
dense regions close to the disk midplane cannot be traced anymore.
Consequently, the disk brightness profile which is very steep in the millimeter range
becomes significantly flatter at shorter wavelengths.
In other words, due to the increasing optical depth in the disk midplane, the relative
contribution of the lower density outer regions and the directly heated -- and therefore warm --
upper disk layers 
(disk photosphere above the effective disk surface at $\tau_{\rm optical}$\about1)
increases with respect to the contribution from the disk midplane.
This effect is amplified by the decrease of the disk scale height towards the star,
resulting in a further decrease of the amount of dust in the upper, optically thin disk layers.
The relative amount of dust that can efficiently contribute to the disk reemission
decreases towards the star compared to the total amount of dust at the same line of sight.
Depending on the optical properties of the dust grains and the particular disk structure, 
the radial intensity profile might thus even decrease,
resulting in a local minimum at the position of the central star.
The amount of dust grains which efficiently contribute to the flux measured
at 1.36\,mm (OVRO), but not at 894\,$\mu$m (SMA) depends sensitively both on 
the radial {\em and} vertical distribution of the disk density and dust grain properties
and thus also on the disk temperature structure.

Based on the WPS03 disk model outlined in Sect.~\ref{s.sect.model}, the predicted
size of the region around the star in which the flux decreases is too narrow
to be spatially resolved by our SMA observations, i.e.,
the quantitative behavior of the radial brightness profile
is different in the simulation and observation for the inner \about80-120\,AU
(see Fig.~\ref{sma.radprof}).
Consequently, the opacity structure, and therefore the quantities
that determine it (outlined above), are different in this region than assumed in the model.
As we assume perfect mixing of the dust throughout the entire disk,
i.e., the dust properties and the gas-to-dust mass ratio are the same
at every point in the disk, this finding is not surprising.
Indeed, detailed theoretical investigations of the planet forming process,
which is expected to take place in this inner region,
predict that these assumptions have to be given up because
of the radial and vertical dependence of the evolution of the dust and disk parameters
(e.g., Nomura \& Nakagawa~2006).
However, a self-consistent model which takes into account
the various processes involved 
(e.g., grain growth and fragmentation, dust sedimentation, radial and vertical mixing; 
see Sect.~\ref{sect.sum})
is beyond the scope of this publication.

A very clear test of our conclusions will be possible with the Atacama Large Millimeter Array (ALMA; Wootten~2003).
Thanks to its significantly higher spatial resolution it will not only be able
to trace the local minimum centered on the stellar position, 
but also detect the brightness decrease in vertical direction, centered
on the disk midplane.
The resulting quadrupolar disk structure can be seen in the predicted ideal 0.3\,mm, 1.36\,mm, and 2.74\,mm maps
(Fig.~4, WPS03).
However, since the optical depth effect is even stronger than predicted, the effect will be even more pronounced
for a given wavelength.

\section{Summary and Conclusions}\label{sect.sum}

We obtained the first spatially resolved submillimeter map of circumstellar disk 
of the prominent Butterfly Star in Taurus. 
We find a good agreement between the observed and the predicted brightness profile
in the outer region of the disk.
However, a discrepancy between the predicted and observed radial flux distribution
was found in the inner region (inside \about80-120\,AU),
showing a decrease of the flux towards the center.
This discrepancy -- a local minimum at the stellar position -- amounts to only \about2$\sigma$.
However, the specific location of the minimum at the disk center and 
the slight symmetry of the corresponding local maxima with respect
to the disk center indicate that this profile is 
an intrinsic feature of the real disk brightness distribution.
Based on radiative transfer simulations and additional observations at 1.36\,mm
we exclude a large inner hole (strongly depleted from small dust grains) as a possible
explanation of this local minimum in the radial brightness distribution.

These observations may provide the basis for a detailed model of the inner structure
of the disk of the Butterfly Star. While earlier modeling lead to conclusion
about grain growth in the disk (WPS03), the new SMA observations provide constraints
for the disk structure and dust grain properties in the inner, potentially
planet-forming region (\about80-120\,AU) vs.\ the outer disk region (\about120-300\,AU). 
It was found that the optical depth in the inner disk region
is higher than predicted by WPS03.
The column density of dust along the line of sight is therefore higher
in the inner disk region than derived from the previous disk model 
in which a perfect mixing of dust and gas throughout the entire disk is assumed.
These observations indicate that there exists a higher dust density and
therefore a higher dust-to-gas mass ratio in the inner disk region.
This conclusion is based on the assumption that the density profile
of the gas phase of the disk can be described by the same approach 
in the inner and outer disk region (Eq.~\ref{dendisk})
and that there exists no discontinuity in the gas density profile.

Since the optical depth effect discussed above is constrained by both 
the radial and vertical disk structure and dust grain properties,
such a model will have to take the predicted grain evolution and its 
dependence on the radial and vertical position in the disk into account.
Furthermore, dust settling and the resulting
increase of the grain-grain interaction probability will result in a further
vertical dependence of the grain size distribution
(e.g., Weidenschilling~1997).
Beside the grain evolution, mixing processes, such as convection
and radial mixing within circumstellar disks have to be considered
since these processes lead to a redistribution of processed, evolved dust grains
to outer, less dense and colder disk regions
(see, e.g., Klahr et al.~1999, Gail~2001). 

Our observations illustrate the high potential of submillimeter observations
for studying circumstellar disks around young (proto)stars. 
Although the spatial resolution
is significantly lower than that aimed for in the case of ALMA in a few years from now,
the tight correlation between the density and temperature structure in the disk
is already able to constrain the radial and vertical disk structure 
and thus the dust grain properties as a function of the distance from the disk midplane,
based on
the spatially resolved radial brightness profile 
obtained with the SMA
in the submillimeter wavelength range.

\acknowledgments

S.W.\ and A.S.\ are supported by the German Research Foundation (DFG) 
through the Emmy Noether grant WO\,857/2.
H.B.\ is supported by the DFG through the Emmy Noether grant BE\,2578.
We wish to thank the anonymous referee for valuable suggestions
concerning the presentation and discussion of our results.

\end{document}